\documentclass[aps,prd,reprint,preprintnumbers,nofootinbib]{revtex4-1}

\usepackage{amsmath}
\usepackage{amsfonts} 
\usepackage{amssymb}
\usepackage{graphicx}
\usepackage{hyperref}
\usepackage{tensor}
\usepackage{siunitx}
\usepackage{float}
\usepackage{bm}
\newcommand{\be}{\begin{equation}}
\newcommand{\ee}{\end{equation}}

\begin{document}

\title{\Large On the particle picture of Emergence}

\author{\large Jarod Hattab}
\author{\large Eran Palti}

\affiliation{\vspace{0.2cm}
Department of Physics, Ben-Gurion University of the Negev, Beer-Sheva 84105, Israel}

\begin{abstract}
The Emergence Proposal is the idea that all kinetic terms for fields in quantum gravity are emergent in the infrared from integrating out towers of states. It predicts that in a supersymmetric string theory context, the tree-level prepotential terms can be recovered precisely by integrating out a tower of non-perturbative states. In this note we present a new perspective, and associated quantitative evidence, for this proposal. We argue that the tree-level kinetic terms arise from integrating out the ultraviolet physics of each of the states in the tower. This ultraviolet physics is associated to extended objects, and cannot be captured by a standard particle Schwinger integral. Instead, we argue that it should be captured by a Schwinger-like integral where the proper time is analytically continued, and a contour is taken around the origin. This maps to certain integral representations for the moduli space periods, and indeed one recovers the tree-level prepotential exactly. This interpretation suggests that the ultraviolet physics which gives the leading contribution to the prepotential is localised on point intersections of the extended objects. We also argue that over special loci in moduli space there can exist a particle picture of the states, and an associated simple particle Schwinger integral, which leads to the full tree-level prepotential. These are loci with special degenerations, such as the singular limit of the resolved conifold.
\vspace{1cm}
\end{abstract}

\maketitle

\section{Introduction}
\label{sec:int}

As formulated in \cite{Palti:2019pca}, the Emergence Proposal states that the dynamics (kinetic terms) for all fields arise from integrating out towers of states down from a scale below the Planck scale. The proposal is based on the ideas, and certain toy-model calculations, in \cite{Harlow:2015lma,Heidenreich:2017sim,Grimm:2018ohb,Heidenreich:2018kpg,Palti:2019pca}, and on the proposal in \cite{Grimm:2018ohb} that the moduli space of type IIB string theory compactified on a Calabi-Yau manifold arises completely from integrating out D3 branes. More precisely, this is the strong version of the proposal, the one advocated in \cite{Grimm:2018ohb,Palti:2019pca}, where there are no dynamical fields in the ultraviolet. A milder statement is that there is unification of strong coupling of all the fields below the Planck scale. The proposal has been studied in some detail within string theory, see \cite{Grimm:2018cpv,Corvilain:2018lgw,Lee:2019xtm,Lee:2019wij,Palti:2020tsy} for a sample of some early work, and  \cite{Marchesano:2022avb,Castellano:2022bvr,vandeHeisteeg:2022btw,Cota:2022maf,Cribiori:2022nke,Marchesano:2022axe,Blumenhagen:2023tev,Castellano:2023qhp,Burgess:2023pnk,Baume:2023msm,Cribiori:2023ffn,Blumenhagen:2023yws,Blumenhagen:2023xmk,Seo:2023xsb,Calderon-Infante:2023ler,DeBiasio:2023hzo,Calderon-Infante:2023uhz, Castellano:2023aum,Castellano:2023jjt,Castellano:2023stg,Marchesano:2023thx,Basile:2023blg,Cota:2023uir,Marchesano:2023thx}, for more recent investigations. 

Strong Emergence makes sharp predictions for the precise values of tree-level kinetic terms: that they must be exactly reproduced from integrating out states. In practice, checking such a prediction is extremely difficult because one needs to know all the relevant states from which the infrared kinetic terms are emergent, and further, know how to integrate them out exactly. Recently, it was noted in \cite{Blumenhagen:2023tev}, that at least the former problem can be addressed in the case of type IIA string theory on the resolved conifold, where the spectrum of BPS states that could contribute to emergence is very simple. It consists of a single D2 brane wrapping the resolution two-cycle, and all its possible D0 brane bound states (as well as free D0 branes). Further, a way to integrate out all the states using a Schwinger integral approach and zeta-function regularization was also proposed, and it was shown to reproduce a part of the tree-level prepotential. These fascinating developments are the motivation for this note. 

In this note, we present a different perspective on the integrating out procedure to the one of \cite{Blumenhagen:2023tev}. We argue that there cannot be a general Schwinger integral calculation for integrating out the towers of states, and that the tree-level kinetic terms should really arise from integrating certain degrees of freedom of extended objects (which in a certain frame are M-theory membranes).
However, we will propose a certain modification of the Schwinger integral, where the Schwinger proper time is complexified (analytically continued), and the integrand is appropriately modified to capture this ultraviolet physics. These complex integrals will then yield the tree-level prepotential (kinetic terms) of the supergravity theory. We will also argue that the leading contribution to the prepotential arises from integrating out point-like degrees of freedom. Further, we will argue that there can exist certain loci in moduli space where there is a particle representation of the relevant ultraviolet physics, and on those loci it is possible to derive the tree-level prepotential from integrating out particles, with a standard Schwinger integral calculation if desired. 

\section{The Particle and Field Theory pictures}
\label{sec:pft}

In general, the setting we would like to consider is type IIA string theory compactified on a Calabi-Yau (CY) threefold of large volume (so that the supergravity approximation holds). The states we would like to integrate out are D2-D0 bound states wrapping two-cycles in the CY. The two-cycle volumes are denoted $t^i$, and are the real parts of the bosonic components of chiral superfields $T^i$. The imaginary parts $b^i$ are periodic axions. So we have
\be
T^i=t^i + i b^i \;.
\ee

The kinetic terms for the massless lower-dimensional fields are controlled by the supergravity prepotential ${\cal F}_0$, which is holomorphic in the $T^i$. In the limit of large $t^i$, the prepotential splits into a polynomial piece $\left.{\cal F}_0\right|^{\mathrm{Poly}}$, which captures what one may refer to as tree-level physics, and a piece exponential in the $T^i$ which we denote $\left.{\cal F}_0\right|^{\mathrm{Inst}}$.

There are two approaches to integrating out the wrapped branes. The first is the field theory picture of Gopakumar and Vafa, introduced in \cite{Gopakumar:1998ii,Gopakumar:1998jq}. In this approach, one integrates out through a Schwinger integral where we give the graviphoton a background expectation value. The idea is that one can take the strong-coupling limit of type IIA such that the D2-D0 states become light and can be treated as elementary particles. Due to supersymmetry, because the dilaton is in the hypermultiplet sector, this limit is under control with respect to the Schwinger calculation. The result, after accounting for the supersymmetric properties of the states, gives the contribution to the supergravity prepotential ${\cal F}_0$ of
\be
\left. {\cal F}_0 \right|_{\mathrm{FT}}^{\mathrm{Inst}} =  \sum_{\beta,n \in \mathbb{Z}} \int_{\epsilon}^{\infty} \frac{ds}{s^3}\alpha_{\beta} e^{-s Z_{\beta,n}} \;.
\label{schwintgen}
\ee 
Here $\beta$ denotes the wrapping vector of the D2 branes, $n$ is the number of bound D0 branes, and $\alpha_{\beta}$ is the degeneracy of BPS states with that wrapping vector. Note that the degeneracy with respect to the D0 charge is always just one. \footnote{Note that we work in units where $M_s=2 \pi g_s$, with $M_s$ the string scale, and $g_s$ the string coupling.} $Z_{\beta,n}$ is the (holomorphic) central charge of the associated BPS state, and is given explicitly (in the large $t^i$ regime) by
\be
Z_{\beta,n} =  2\pi \left(\beta \cdot T + in\right) \;.
\ee

A crucial role is played by the integral cutoff $\epsilon$. Since $s$ can be interpreted as the Schwinger proper time, the $s \rightarrow 0$ limit is a short time limit, and so $\epsilon$ is an ultraviolet cutoff. In the field theory picture, we cannot trust the $\epsilon \rightarrow 0$ limit because at such high energy scales the D2-D0 states cannot be treated as particle excitations and instead their internal structure is probed. So we would need to quantize their excitations to understand this regime fully. 

In fact, the integral (\ref{schwintgen}) is independent of the value of $\epsilon$, as long as it is finite $\epsilon > 0$. The ultraviolet physics contribution to the prepotential is therefore captured fully by the pole at $\epsilon=0$. This discontinuity is most easily seen in the particle picture to which we turn. 

\begin{figure}
\centering
 \includegraphics[width=0.45\textwidth]{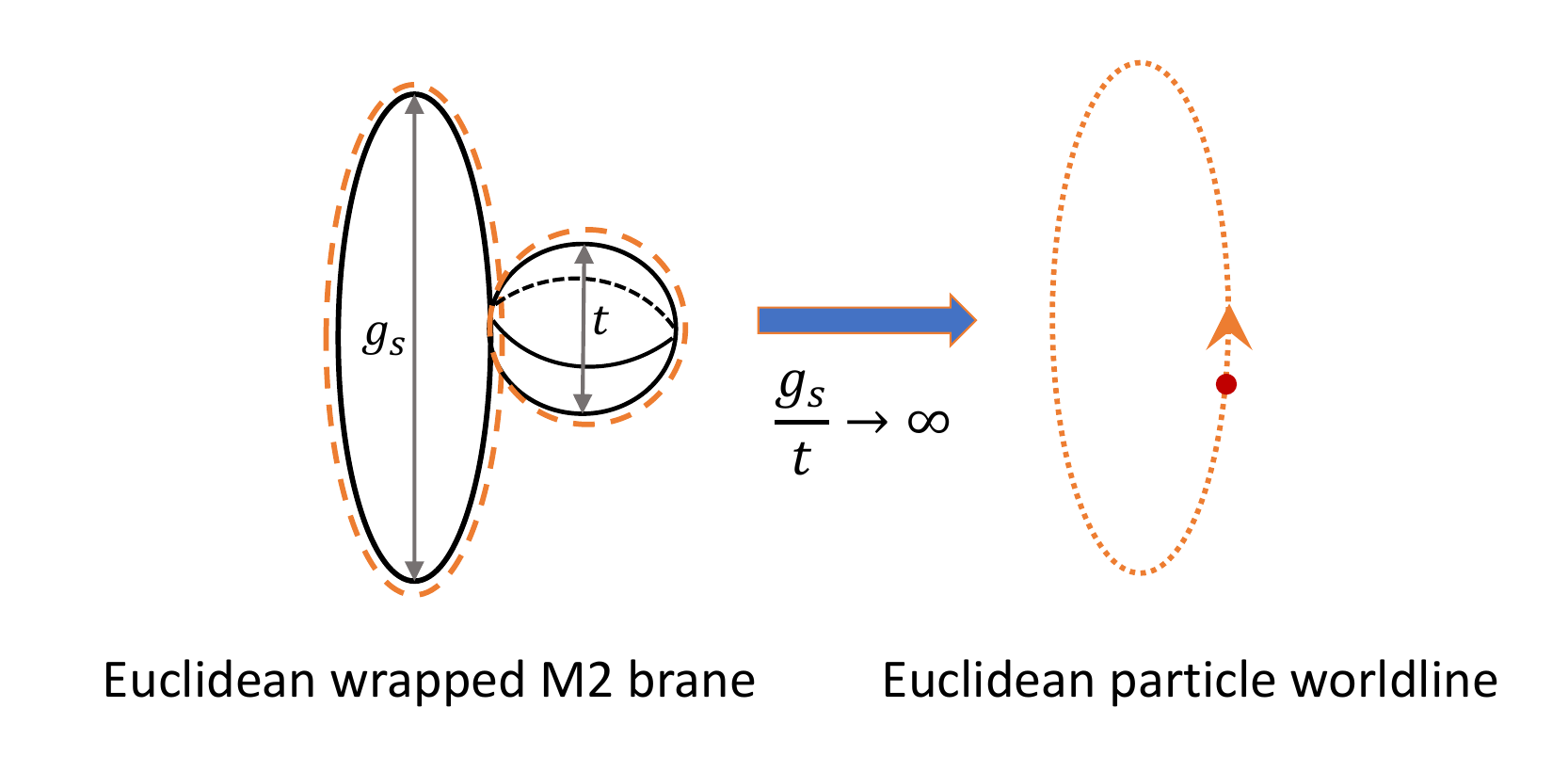}
\caption{Figure illustrating the particle picture where one integrates out the small two-cycle and treats the M2 instanton as the worldline of a Euclidean particle. The worldline length is associated to the string coupling $g_s$, while the two-cycle is controlled by the Kahler modulus $t$.}
\label{fig:partpic}
\end{figure}

The particle picture was introduced in detail in the paper \cite{Dedushenko:2014nya} (although was very much part of the ideas in  \cite{Gopakumar:1998ii,Gopakumar:1998jq}). Here one thinks more directly in terms of the worldsheet instanton contributions to the prepotential. Going to strong coupling, the M-theory picture is that of Euclidean M2 branes wrapping the two-cycles in the CY as well as the M-theory circle. In the limit where the M-theory circle becomes large, so the IIA strong coupling limit, we can integrate over the two-cycles in the CY, and obtain a Euclidean worldline action for a particle. This is the regime where the particle picture is valid, and is illustrated in Figure \ref{fig:partpic}.

The worldline path integral, including the 1-loop determinant, can then be done to yield the resulting instanton contributions. The result is\footnote{In \cite{Dedushenko:2014nya}, the calculation is only for a single brane. We present here a straightforward generalisation accounting for degeneracy.}
\be
\left. {\cal F}_0 \right|_{\mathrm{P}}^{\mathrm{Inst}} = \sum_{\beta,k \geq 1} \frac{1}{k^3} \alpha_{\beta} e^{-2\pi k \beta \cdot T} \;. 
\label{partigen}
\ee
Here $k$ is the winding number of the M2 brane on the M-theory circle. It is important that for $t^i>0$ only the $k \geq 1$ contribution can be reliably calculated because the $k=0$ state is an M2 brane which is not wrapping the M-theory circle at all. In that case, there is no hierarchy between the size of the different dimensions that it is wrapping and so we cannot integrate out the CY two-cycle and write an effective particle worldline. Instead, the $k=0$ state is sensitive to the ultraviolet details of the M2 brane, and requires quantizing it to fully understand.

The particle picture (\ref{partigen}) and the field theory picture (\ref{schwintgen}) are related by a Poisson resummation
\be
\sum_{k \in \mathbb {Z}} \delta\left(s-k\right) = \sum_{n \in \mathbb {Z}} e^{2 \pi i n s} \;,
\label{poiresum}
\ee
which gives the desired relation $\left. {\cal F}_0 \right|_{\mathrm{P}}^{\mathrm{Inst}} = \left. {\cal F}_0 \right|_{\mathrm{FT}}^{\mathrm{Inst}}$. 
Intuitively, one can think of this a type of T-duality along the time direction, which interchanges an instanton (particle worldline picture) with a particle (field theory Schwinger picture). 

By using (\ref{poiresum}), we see that indeed the integral (\ref{schwintgen}) is independent of $\epsilon$ for any $1>\epsilon >0$, and this integral cutoff corresponds to dropping the zero winding mode $k=0$ in the particle picture. So that in both pictures we are dropping the ultraviolet physics. 

The question of how to recover the tree-level polynomial piece $\left.{\cal F}_0\right|^{\mathrm{Poly}}$ becomes clearer: if it can be recovered, it must be associated to the ultraviolet physics, the $\epsilon=0$ pole in the Schwinger integral, or the dual (unwrapped) $k=0$ membrane. But this is precisely the physics over which we have no control, and no particle description.

\section{A particle approach to the resolved conifold}
\label{sec:rescon}

In \cite{Blumenhagen:2023tev} a calculation was presented of integrating out wrapped D2-D0 states for type IIA string theory on the resolved conifold. The resolved conifold has a single (compact) 2-cycle, parameterised by $T$. The prepotential for this setting was proposed in \cite{Gopakumar:1998ki} to read
\begin{eqnarray}
{\cal F}_0 = -\frac{\left(2\pi i\right)^3}{12} \Bigg[iT^3 + \frac{3}{2}\left(1+4m\right)T^2  - \frac{i}{2} T \Bigg]\nonumber \\- \zeta(3) + \mathrm{Li}_3\left(e^{-2\pi T} \right) \;,
\label{f0coni}
\end{eqnarray}
where $m$ is an arbitrary integer.

The resolved conifold has a particularly simple spectrum of BPS states consisting of a single D2 brane and its D0 bound states. It is a very simple example that will serve to guide us throughout this investigation. 

The field-theory (\ref{schwintgen}) and particle integrals (\ref{partigen}) for the spectrum of states of the resolved conifold are
\be
\sum_{n \in \mathbb{Z}} \int_{\epsilon}^{\infty} \frac{ds}{s^3} e^{-s 2\pi \left(T+in \right)} = \sum_{k \geq 1} \frac{1}{k^3} e^{-2\pi k T} = \mathrm{Li}_3\left(e^{-2\pi T} \right)\;.
\label{intrescon}
\ee 
As expected, the expressions (\ref{intrescon}) reproduce the exponential terms in the prepotential. 

The question of interest for us in this note is whether there is an integrating out calculation that can also reproduce the polynomial piece in the prepotential (\ref{f0coni}). In \cite{Blumenhagen:2023tev} such a calculation was performed, which reproduced the cubic piece in $T$, though not the linear and quadratic ones. We will take a different approach to the calculation, but it could be that there are relations between the approaches.

Our starting point is to just go ahead and try to recover the tree-level prepotential from a standard particle Schwinger integral, but now with $\epsilon=0$. We proceed by performing the Poisson resummation first. One can consider the Schwinger integral (\ref{schwintgen}) after the sum over $n$ and with $\epsilon=0$, 
\be
\sum_{k} \int_{0}^{\infty} \frac{ds}{s^3}\delta\left(s-k\right) e^{-2\pi s T} \;.
\label{schwintgenksum}
\ee 
Our claim is that the polynomial piece is somehow associated with the pole at $s=0$, so that (\ref{schwintgenksum}) captures the full ${\cal F}_0$. We can note that the $T^3$ part of the polynomial piece is independent of this pole, so we can write
\begin{eqnarray}
\frac{\partial^3}{\partial T^3}{\cal F}_0  &=&  -\left(2\pi\right)^3\sum_{k} \int_{0}^{\infty} ds \;\delta\left(s-k\right) e^{-2\pi s T} \\ \nonumber 
&=& -\frac{\left(2\pi\right)^3}{2} -\left(2\pi\right)^3\sum_{k} \int_{\epsilon}^{\infty} ds \;\delta\left(s-k\right) e^{-2\pi s T} \;.
\label{naivparp}
\end{eqnarray}
The first term indeed reproduces the correct $T^3$ dependence in the prepotential, as in (\ref{f0coni}), and the second term is just the contribution from the instantons. 

So it seems that a naive particle picture approach can reproduce at least the $T^3$ piece. However, we know that in doing this we utilized the Schwinger integral in a regime where it is not trustable, at $s=0$. The reason that it worked nonetheless is discussed in section \ref{sec:physint}.

\section{An ultraviolet-modified Schwinger integral}
\label{sec:schres}

On the one hand, it seems that some of the polynomial physics of the prepotential is indeed captured by the pole of the Schwinger integral at $s=0$. On the other hand, we know that the physics must be modified in that region, and that the particle pictures of the Schwinger integral is not trustable. We therefore propose that some ultraviolet modification of the Schwinger integral does capture the appropriate ultraviolet physics. 

To guide us towards such a modification we note that the physics we are after is associated to a pole, and such poles arise most naturally in complex integration. We therefore propose that the Schwinger integral should be analytically continued with a complex time parameter $z$. We take
\be
s \rightarrow z \;,
\ee
and we integrate around a contour surrounding the origin of the complex plane $C_0$.
So we first rewrite the $k=0$ part of the particle integral as
\be
\int_{0}^{\infty} \frac{ds}{s^3}\delta\left(s\right) e^{-2\pi s T} \rightarrow \frac12 \oint_{C_0} \frac{dz}{z^3} \frac{1}{2\pi iz} e^{-2\pi z T}\;.
\label{schwintgenk0ana}
\ee 
However, we know that we must also appropriately modify the integral around the $z=0$ region. This modification captures the ultraviolet physics, and we consider it in two different settings below.

\subsection{The resolved conifold}
\label{sec:rescond}

In the case of the resolved conifold the appropriate modification around $z=0$ for the Schwinger integral is
\begin{eqnarray}
\left. {\cal F}_0\left(T\right) \right|^{\mathrm{Poly}}_{m=0} &=& \frac12 \oint_{C_0} \frac{dz}{z^3} \frac{e^{-2\pi z T}}{1-e^{-2\pi i z}}  
\label{fconipolyuvmodz}
\\ \nonumber
&=&-\frac{\left(2\pi i\right)^3}{12} \Bigg[iT^3 + \frac{3}{2}T^2  - \frac{i}{2} T \Bigg]
\end{eqnarray}
This is a contour integral representation of the Hurwitz zeta function. It reproduces the full polynomial part of the prepotential (\ref{f0coni}).\footnote{Note that the $\zeta(3)$ piece is not reproduced here. This is because it arises from integrating out pure D0 states \cite{Gopakumar:1998ii,Blumenhagen:2023tev}.} 

The restriction to $m=0$ is simply a choice for the origin of the periodic $b$ coordinate, which is periodic under $b \rightarrow b+1$.

An important point is that (\ref{fconipolyuvmodz}) also has poles at positive integer values of $z$. 
So if we change our integration contour to not only go around the origin, but cover the whole positive real line, then we will pick up those poles too. These are then the instanton contributions. It turns out that this does not fully work because of the factor of half in front of the pole at $z=0$, so we get only half the instanton contibutions this way. This is related to the fact that the resolved conifold has a self-intersection number of a half. In section \ref{sec:compcy}, we present a proper Calabi-Yau example, where we see that indeed the instantons and tree-level pieces are just different poles of the same integrand.

This understanding, and ignoring the factor of half issue, explains also how the ultraviolet modification can be determined from the infrared knowledge (the instanton terms). One has to write the sum over instanton terms as a single holomorphic function, such that the instanton contributions come from the positive integer poles of it. The tree-level contribution is then just the zero pole of that function.



\subsection{Compact one-parameter Calabi-Yau}
\label{sec:compcy}

Our proposal is that the ultraviolet physics of integrating out the states can be captured by a Schwinger-like complex integral around the origin. For the conifold, the appropriate ultraviolet modification was (\ref{fconipolyuvmodz}). In this section we consider a more complicated setup which is a proper compact Calabi-Yau. We consider the bi-cubic $\mathbb{P}^5[3,3]$ which was studied in detail in \cite{Joshi:2019nzi,Palti:2021ubp} (see also \cite{Bastian:2023shf} for a general context study). More precisely, we can consider the mirror to it, which has one Kahler parameter $T$. However, our approach is general, and can equally be understood on the type IIB side with a complex-structure field, indeed that is the more reliable and precise setting to consider since it captures the full moduli space.

There are two parts to deriving the prepotential from integrating out the states: the first is how to integrate out a single state, with some appropriately modified integral in the ultraviolet, and the second is how to sum over the states. We know that the sum over the D0 parts of the D2-D0 states can be exchanged for a delta function, as in (\ref{poiresum}). This delta function is then exchanged for a pole in the complex formulation of the integral. The sum over the D2 parts of the states is more complicated. As discussed in section \ref{sec:rescond}, this sum is encoded generally as a sum over poles of a single meromorphic function.\footnote{More precisely, the sums over the D2 wrapping and the D0 charge are re-summed such that the poles are at integers which are the product of the D2 and D0 charges.} The continuation of this function to the origin then gives the required ultraviolet modification to yield the tree-level prepotential. Therefore, the two aspects discussed at the start of this paragraph are only one: once you know how to write the sum over states as a function, you also know the ultraviolet modification.

We are after a complex Schwinger type integral in which already the sum over states has been performed, and which recovers the tree-level polynomial prepotential through a pole at the origin. Such integrals have been known for a long time, they are the integral representations of the periods. Our results are a simple manipulation of these integrals into an expression for the prepotential. For this example Calabi-Yau case, we can utilise the results in \cite{Joshi:2019nzi} for integral representations of the periods to find an appropriate Schwinger-type integral for the prepotential. We derive
\begin{eqnarray}
\left.\mathcal{F}_0\left(T\right)\right|^{\mathrm Poly} &=&
\frac12 \oint_{C_0}\biggl[\left(3 + i \sqrt{3}\right)\frac{\Gamma\left(\frac13+z\right)(-1)^{-z}}{\Gamma\left(\frac23-z\right)\Gamma\left(\frac13-z\right)^2} \nonumber  \\
\label{polyint}
&+&\left(-3 + i \sqrt{3}\right)\frac{\Gamma\left(\frac23+z\right)(-1)^{-z}}{\Gamma\left(\frac13 -z\right)\Gamma\left(\frac23 -z\right)^2} \\
\nonumber 
&-& \frac{(3i + 9\sqrt{3})}{ 2 \pi}\frac{\Gamma\left(\frac13+z\right)^2}{\Gamma\left(\frac13 -z\right)^2}
\nonumber \\
&+& \frac{ \left(9\sqrt{3}-3i\right)}{ 2\pi}\frac{\Gamma\left(\frac23+z\right)^2}{\Gamma\left(\frac23 -z\right)^2}\biggr]\frac{\Gamma\left(-z\right)^4 3^{6z}e^{-2\pi zT} }{ 2\pi i z }
 \nonumber\\
&=& -\frac{\left(2\pi i\right)^3}{4}\Big[6iT^3-T^2+9iT \Big] + \mathrm{const.} \nonumber \;.
\end{eqnarray}
We therefore indeed recover the tree-level prepotential. 

\begin{figure}
\centering
 \includegraphics[width=0.45\textwidth]{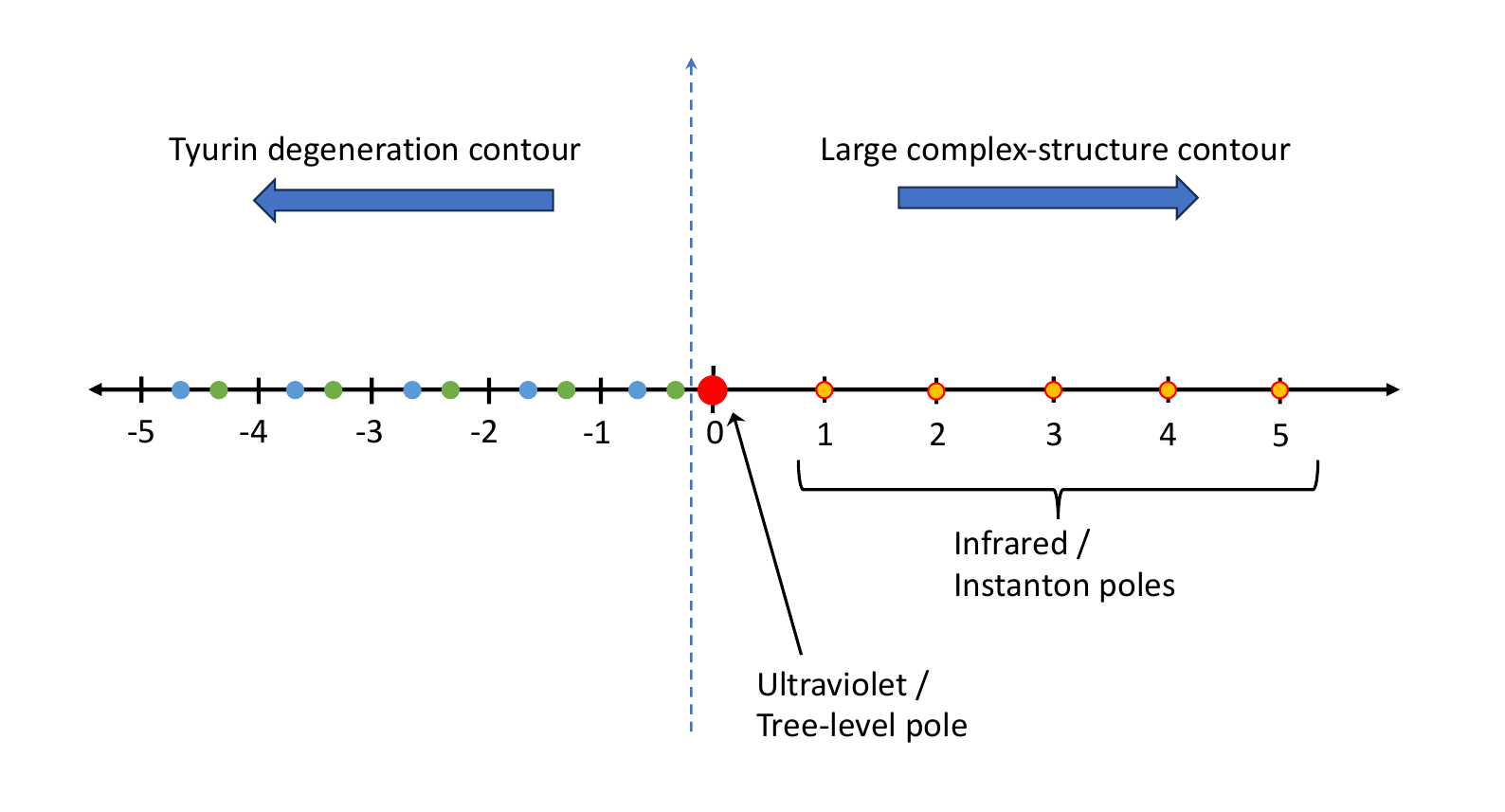}
\caption{Figure illustrating pole structure and the associated physics. The zero pole gives rise to the tree-level prepotential at large complex-structure (mirror to large volume), which from the Schwinger perspective is the ultraviolet physics. The other, infrared poles, give the instanton contributions.}
\label{fig:poles}
\end{figure}

Although our primary interests lie in the pole at $z=0$, which is capturing the tree-level prepotential, and from the Schwinger integral perspective is associated to the ultraviolet. We can also recover the instanton terms, or the Schwinger infrared physics, from the other poles. Specifically, there are two important integration contours with an infinite number of poles, which capture the two infinite distance loci in the moduli space. The contour which includes the $z=0$ pole, and all the positive integer poles gives the prepotential at large complex-structure (mirror to large volume), while the contour which covers all the negative poles gives the prepotential at the other infinite distance locus (Tyurin degeneration). This is illustrated in figure \ref{fig:poles}. The fact that the integral also captures the instanton terms, which we know account for the D2 brane multiplicities, shows that the sum over the states has already been performed and implemented into the integral.\footnote{There is a subtlety here because the instantons are not given exactly by the integral (\ref{polyint}). More precisely, they are given in terms of the periods, and the map between the periods and the prepotential itself involves exponentially small terms in $T$. This is only a technical issue, conceptually one can just think of the integral for the periods rather than the prepotential, which is a single integral that captures the tree-level and instanton terms together.} 

\subsection{The physics interpretation of the modified Schwinger integral}
\label{sec:physint}

The modified Schwinger integral (\ref{polyint}), which reproduces the tree-level prepotential, can shed some light on the physics which is being integrated out. We are concerned with the pole around $z=0$, and so it is informative to expand the integral about that pole
\begin{eqnarray}
\left.\mathcal{F}_0\left(T\right)\right|^{\mathrm Poly} &=&
\frac12 \oint_{C_0}\biggl[18z + {\cal O}\left(z^2\right)\biggr]\frac{\Gamma\left(-z\right)^4 3^{6z}e^{-2\pi zT} }{ 2\pi i z }
\nonumber \\
&=& \left(2\kappa\right) \frac12 \oint_{C_0} \frac{1}{z^3}\frac{1}{2\pi i z} e^{-2\pi z T} + ...\;.
\label{selfinint}
\end{eqnarray}
Here $\kappa=9$ is the self intersection number of the cycle in the Calabi-Yau. 
The approximation in (\ref{selfinint}) is only keeping the contribution to the $T^3$ piece, which is sufficient for our discussion.

\begin{figure}
\centering
 \includegraphics[width=0.45\textwidth]{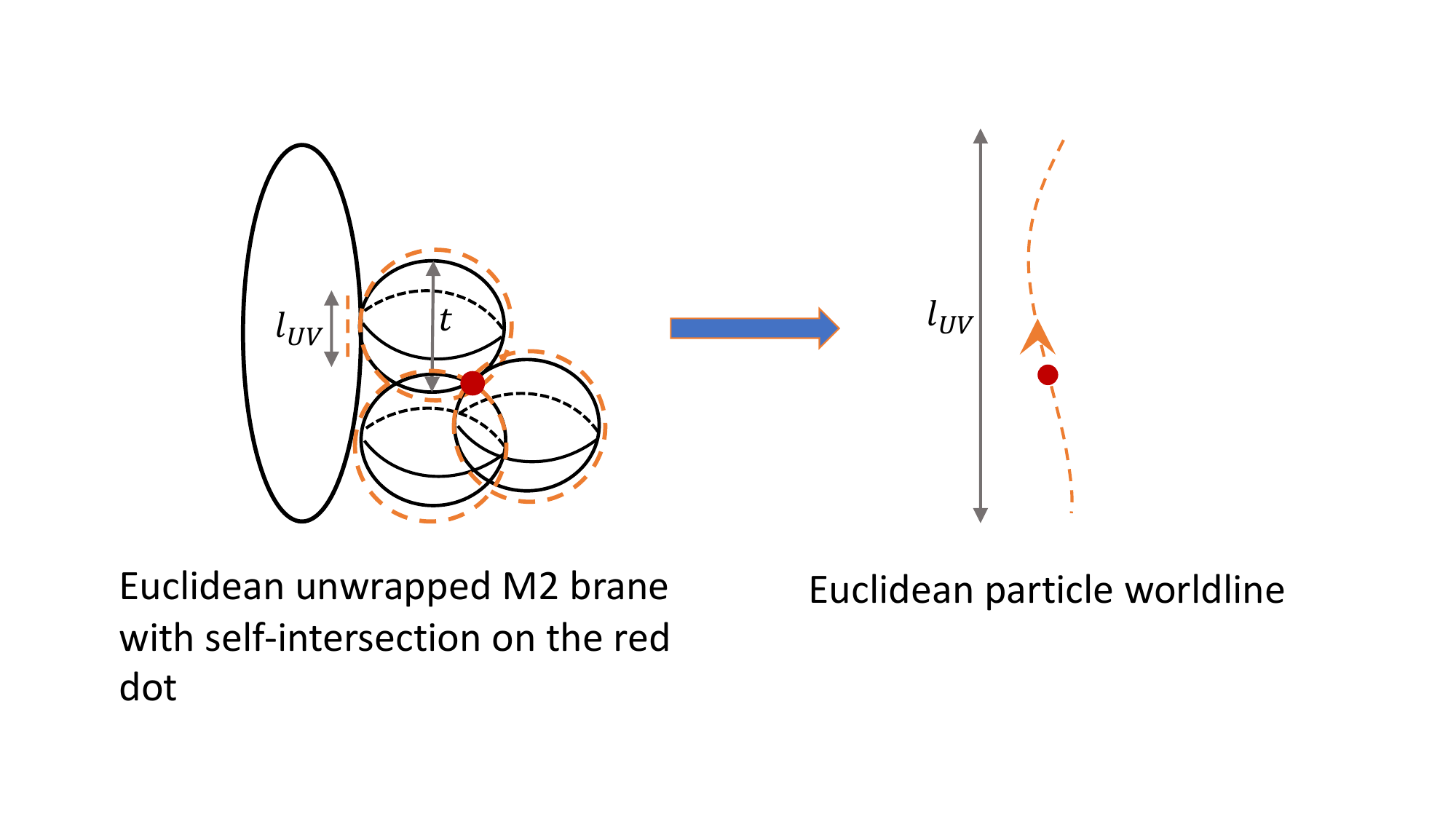}
\caption{An illustration of how degrees of freedom which are localised at self-intersection points can be treated as a particle world-line picture even when the membrane is not wrapping the M-theory circle and only propagating a small ultraviolet scale $l_{UV}$.}
\label{fig:partint}
\end{figure}

Comparing (\ref{selfinint}) and (\ref{schwintgenk0ana}) we see that this is like $2\kappa$ copies of the simple analytic continuation of a single particle contribution. The factor of $\kappa$, which counts the point-like self-intersections of the branes is telling us that the contribution to the pole is coming from integrating out degree of freedom associated to the intersection points. Since these are localised, they behave like particles, and this is why we can recover the $T^3$ piece from a naive particle approach as in (\ref{naivparp}). This is illustrated in figure \ref{fig:partint}. Indeed, this make sense due to our interpretation of $z$ as an analytic continuation of Schwinger proper time, and so $z=0$ being the zero time, or ultraviolet, limit of the physics: only point-like contributions can exist at zero Schwinger time.\footnote{The factor of $2$ in $2\kappa$ is likely related to the phenomenon in section \ref{sec:parrescon}, where we see that the particle picture involves pairs of oppositely charged particles.}

More generally, the integral representation (\ref{polyint}) is showing us that the contribution of integrating out degrees of freedom to the prepotential is ordered by their size scales, or Schwinger proper time length. Contributions at scales smaller that the two-cycle volumes no longer see the whole wrapped object and are associated to physics localised within the brane, and are polynomial. After those, we reach scales larger than the cycle volumes which now see the whole brane. These scales then lead to the familiar exponential suppression in the mass, and the higher the wrapping number/length, the stronger the suppression.

The idea that the contribution to the polynomial terms is coming from localised degrees of freedom inside the branes also matches expectations from supersymmetry. Specifically, we expect to restore higher supersymmetry in the ultraviolet, apart from at the localised points. Higher supersymmetry implies that the prepotential does not receive corrections. So only the localised parts which break this higher supersymmetry can contribute corrections. This can also explain how emergence can work for higher supersymmetric settings: it is not integrating out the half-BPS whole branes, but the localised degrees of freedom inside them which break more supersymmetry.

There is another mystery which can be understood from a Schwinger proper time interpretation of the integral (\ref{polyint}): in the infrared we have a huge degeneracy of D2/M2 branes for each cycle homology class, exponentially growing with wrapping number, but the contribution to the tree-level prepotential seems to have no such degeneracy. In other words, we would naively expect to get a $2\kappa$ contribution from each wrapped D2/M2 brane, not just a single total contribution. We believe that the explanation for this is that one must account for the scale at which the contribution is evaluated. The spectrum of strongly-bound states depends on the scale one works at, and it can be that at different scales what was a strongly-bound state becomes a weakly-interacting state of multiple branes. In such cases, that state is already accounted for by integrating out the constituents. When we count the branes in the infrared, we only integrate at each scale the strongly-bound states. What our picture suggests is that at scales below that of the wrapped branes, so the ones relevant for the tree-level prepotential, all the relevant states (which are the localised degrees of freedom in the sum over cycle wrappings but zero M-theory circle winding) are weakly-interacting states of the one set of degrees of freedom localised at the intersections.\footnote{We expect that this is because, like all three-dimensional theories, M2 branes are strongly interacting in the infrared, and weakly-interacting in the ultraviolet.}


\section{A particle limit of the Conifold}
\label{sec:parrescon}

\begin{figure}
\centering
 \includegraphics[width=0.45\textwidth]{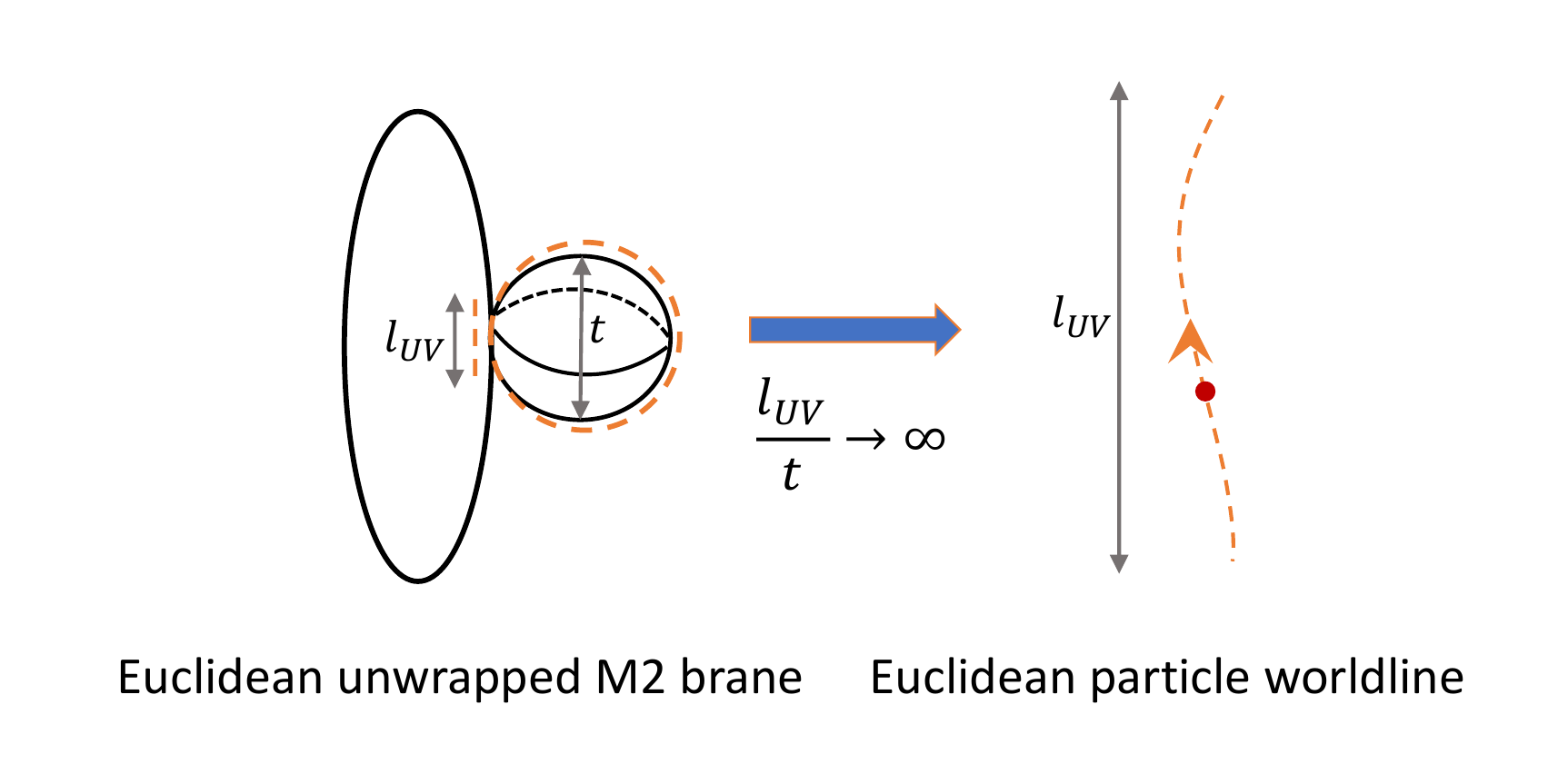}
\caption{Figure illustrating the particle picture for unwrapped $k=0$ membrane states. Since the membrane is not wrapping the M-theory circle, it only propagates a small ultraviolet scale $l_{UV}$ (such as the Planck scale). This means for any finite $t>0$ there is no hierarchy of scales and no worldline picture. However, such a picture may become valid in the limit $t \rightarrow 0$.}
\label{fig:partpicnw}
\end{figure}

We have seen that for a finite size two-cycle $t>0$ we cannot trust a particle picture for the integrating out procedure in the ultraviolet. There is no effective particle world-line description of the type illustrated in figure \ref{fig:partpic}.\footnote{Of course, as discussed in section \ref{sec:physint}, a certain part, associated to localised degrees of freedom, does have particle-like behaviour.} For this reason, we proposed a modification to the Schwinger integral in the ultraviolet. However, it is possible to imagine that certain loci in moduli space exist where a particle picture can be used. Specifically, consider the conifold setting and take the $t \rightarrow 0$ limit. In this limit, the two-cycle shrinks to zero volume. In that case, we may actually be able to use the particle picture, because the worldline of the particle, which is given by some ultraviolet scale $l_{UV}$, is still infinitely longer than the two-cycle size. This is illustrated in Figure \ref{fig:partpicnw}. 

This suggests that a particle picture may be valid on the $t=0$ locus. Although the resolved conifold setting is a supergravity setting with large $t>0$, we will consider evaluating it formally at $t=0$.\footnote{Actually, the derivation of the prepotential in \cite{Gopakumar:1998ki} is a topological string calculation which is insensitive to curvature corrections and so may be reliably used.} In this case, the polynomial part of the prepotential reads
\be
\left. {\cal F}_0 \right|^{\mathrm{Poly}}_{t=0,m=0} = -\frac{\left(2\pi i\right)^3}{12} B_3\left(b\right) \;,
\label{t0prepo}
\ee
where $B_3$ is the third Bernoulli polynomial.\footnote{To be precise, $B_3\left(b\right)$ does not include the $\zeta(3)$ piece (which is associated to pure D0 states).}
Now it is possible to write
\be
\left. {\cal F}_0 \right|^{\mathrm{Poly}}_{t=0,m=0} = \frac12 \sum_{k \neq 0} \frac{1}{k^3} e^{2\pi i k b} \;.
\label{foparconlim}
\ee
Comparing with (\ref{partigen}), this indeed looks like it has a particle picture interpretation. Recall that our proposal is that the tree-level, or polynomial, part of the prepotential comes from the $k=0$ sector, so with no net winding. In (\ref{foparconlim}) we see a sum which is symmetric over winding, so can be thought of as pairs of particles with no net winding. There is also a related overall factor of a half for these pairs. It therefore is naturally interpreted as some kind of a particle picture resolution of the ultraviolet zero net winding sector of the M2 branes.  

An important aspect of (\ref{foparconlim}) is that it does not manifestly transform under $b \rightarrow b+1$, while $B_3$ does transform. This is related to the parameter $m$ in the prepotential. It corresponds to (twice) the transformation property of the Bernoulli polynomial
\be
B_3\left(b+1\right)=B_3\left(b\right) + 3 b^2 \;.
\ee 
The expression (\ref{foparconlim}) is only valid in the range $0 \leq b \leq 1$. So it gives a particle picture resolution within a single monodromy orbit. Indeed, any particle picture of the tree-level physics can only hold in a single monodromy orbit, because (\ref{partigen}) is manifestly invariant under monodromies, while the polynomial part of the prepotential manifestly does transform.  

We therefore find evidence that the polynomial prepotential can be reproduced precisely from integrating out the states, using the particle picture, on the locus $t=0$. If we want to extend this to all $T$, we need to first use the integral representation for the Bernoulli polynomial, and then analytically continue $b \rightarrow -iT$. This yields precisely (\ref{fconipolyuvmodz}). 

\section{Discussion}
\label{sec:disc}

In this note we examined the idea of integrating out states in string theory to recover the tree-level, or polynomial, part of the prepotential. We argued that this should be thought of as ultraviolet physics, and therefore cannot in general have a particle picture interpretation. We proposed that we should modify the Schwinger integral by analytically continuing into complex Schwinger proper time, and modifying the integral in the ultraviolet. This gave a sort of Schwinger integral representation for the tree-level prepotential, and suggests that indeed it arises from integrating out the ultraviolet physics. We note that these integral representations for the tree-level prepotential were known for a long time. More precisely, they were formulated as integral representations for the periods. Our results are technically only a simple manipulation of these integrals, and conceptually, a reinterpretation of them as a type of Schwinger integral. This also yields a general prescription for calculating the ultraviolet modification: it is simply to write the sum over instantons as a single meromorphic function, and extend that function to the origin.

This reinterpretation also sheds light on which physics of the ultraviolet is being integrated out. In particular, the leading contribution to the prepotential was argued to arise from degrees of freedom localised within the branes, at their intersections. This actually allows for the leading piece to be treated effectively as a particle contribution.

We also argued that on certain loci in moduli space, such as the singular limit of the resolved conifold, a particle picture of the states being integrated out for the full tree-level prepotential does exist.

It is important to emphasise that the physics which is being integrated out is the ultraviolet physics of the full tower of states which become light at infinite distance. It is only after summing over this tower that it can be interpreted as fixed zero winding in the particle picture. So we are really integrating out the light tower of states, and the infrared physics of each state gives a contribution to the instantons, and the ultraviolet physics of each state gives a contribution to the tree-level prepotential. More precisely, in the D2/D0 picture, the D0 tower is integrated out completely and replaced with a pole in the Schwinger integral. The D2 tower instead is argued to not contribute to the tree-level prepotential because in the ultraviolet the relevant localised degrees of freedom in the tower are weakly-interacting states of a single set of localised degrees of freedom. 

Also, we are looking at only the infrared part of the physics which has been integrated out, so its effects on the massless spectrum. This is why there is some chance to capture it at all, otherwise we would be needing to capture the full microscopic physics of extended objects. This is perfectly in line with the understanding of emergence as an infrared duality \cite{Harlow:2015lma,Grimm:2018ohb,Palti:2019pca}: the light fields are dual to certain degrees of freedom of the tower of states, they are the same physics described in different variables. 

Finally, it is perhaps worth emphasising that, since emergence is an infrared duality, it can never occur by integrating out states that are fundamental in the same frame as the emergent fields. That would be counting the same physics twice. So Kaluza-Klein photons cannot arise from integrating out Kaluza-Klein modes of fundamental fields, and emergent string limits (using the language of \cite{Lee:2019wij}) cannot emerge from integrating out string oscillator modes. This also makes it clear that the emergence scale for limits where the lightest tower is that of string oscillators, cannot be the string scale. It is sometimes said that the Species scale for string towers is the string scale. If that is how it is defined, then certainly the scale of emergence is not the species scale in general, but should be labelled as an independent Emergence scale. The Emergence scale is the scale where gravitational interactions become strong, whether that be Einstein gravity or string gravity, and string theory is weakly coupled at the string scale. For emergent string limits, the Emergence scale lies between the string scale and the Planck scale, and is associated to the tension of non-perturbative states, typically some branes.

\vspace{0.1cm}
{\bf Acknowledgements}
\noindent
We would like to thank Ralph Blumenhagen. Niccolo Cribiori, Max Wiesner and Timo Weigand for useful discussions. 
 The work of JH and EP is supported by the Israel Science Foundation (grant No. 741/20) and by the German Research Foundation through a German-Israeli Project Cooperation (DIP) grant ``Holography and the Swampland". The work of EP is supported by the Israel planning and budgeting committee grant for supporting theoretical high energy physics.
\appendix

\bibliography{Higuchi}

\end{document}